\documentstyle[prl,aps,psfig,multicol]{revtex}
\def \beq{\begin{equation}}
\def \eeq{\end{equation}}
\begin{document}
\draft
\preprint{\hbox{\begin{tabular}{c}
                hep-lat/9906018 \\
                TIFR/TH/99-29 \\
\end{tabular}}}
\title{Probing the quark-gluon plasma with a new Fermionic correlator }
\author{R.\ V.\ Gavai\cite{RVG} and Sourendu Gupta\cite{SG}}
\address{Department of Theoretical Physics, Tata Institute of Fundamental
         Research,\\ Homi Bhabha Road, Mumbai 400005, India.}
\maketitle
\begin{abstract}
We present the first measurement of a new correlation function of Fermion
bilinears in finite temperature QCD with and without dynamical quarks in a 
quantum number channel in which non-trivial correlations are known to
be present for purely gluonic operators. We find that the Fermion 
correlator vanishes for $T \ge 3T_c/2$, in agreement with the
expectation for weakly interacting quarks in a quark-gluon plasma.
\end{abstract}
\pacs{11.15.Ha, 12.38.Mh \hfill hep-lat/9906018, TIFR/TH/99-29}
\begin{multicols}{2}

The Relativistic Heavy Ion Collider (RHIC) in BNL, New York and the Large
Hadron Collider (LHC) in CERN, Geneva may yield a new state of matter,
called Quark-Gluon Plasma (QGP), which could have existed in our universe
a few microseconds after the Big Bang. It is a theoretically challenging
task to deduce from first principles as many properties of the plasma as
possible. Such a program may help in devising clear and unique signals
of QGP.

Lattice simulations of Quantum Chromo Dynamics (QCD) have provided a 
robust approach based on first principles towards this end. Such
simulations of field theories in equilibrium at finite temperature
($T$) use a discretisation of the Euclidean formulation for partition
functions---
\begin{equation}
   Z(\beta)\;=\;\int{\cal D}\phi\exp\left[-\int_0^{1/T} dt\int d^3x
      {\cal L}(\phi)\right],
\label{int.part}\end{equation}
where $\phi$ is a generic field, $\cal L$ the Lagrangian density, and
the Euclidean ``time'' runs from 0 to $1/T$. The path integral is over
field configurations which are periodic (anti-periodic) in Euclidean time
for the Bosonic gluon (Fermionic quark) fields. Due to a lack of symmetry
between the space and Euclidean time directions in eq.\ (\ref{int.part}),
this problem has only a subgroup of the full 4-dimensional rotational
symmetry of the $T=0$ Euclidean theory. Since the partition function above
contains equal weights for all configurations which are related by
these symmetries, only those operators which transform as scalars under this
reduced symmetry group have non-vanishing expectation values.

For the lattice discretised problem the symmetry groups reduce to discrete
subgroups of the continuum symmetry groups. It is useful to
write the partition function of eq.\ (\ref{int.part}) as the trace of
the transfer matrix in one of the spatial directions. Correlation
functions along that direction can then be classified by the
irreducible representations (irreps) of the symmetry group of the
transfer matrix. Unlike operator expectation values, correlation
functions are generally non-vanishing in all representations--- not just
the scalar. At $T=0$, the symmetry group of the transfer matrix for QCD
using the staggered Fermions for quarks has been studied extensively,
and the representations of corresponding Fermion bilinear correlation
functions are well known \cite{mesons}. The symmetries and representations
of screening correlation functions at finite temperature have been worked
out recently \cite{fermit}.

The main point of this last analysis is that the symmetry group
of the $T>0$ transfer matrix is smaller than that of the $T=0$
transfer matrix. As a result, the $T=0$ irreps reduce further at
finite temperature. All correlation functions block diagonalise under
the isometry group of the spatial slice of the thermal lattice, the
dihedral group $D_4^h$. As an example, a correlation function in, say,
the $z$-direction of any vector (V) or pseudo-vector (PV) operator,
$(V_x,V_y,V_t)$, in the $T=0$ theory breaks up into two scalar ($A_1^+$)
irreps of $D_4^h$, the components $V_t$ and $V_x+V_y$, and a $B_1^+$
irrep $V_x-V_y$. This happens for gluonic Wilson-loop operators such as
plaquettes, and also for the quark bilinears operators  \cite{fermit}.

The plaquette operators, restricted to a $z$-slice, transform as a
PV set at $T=0$ and provide a good example of this reduction.
The combinations $P_{xy}$ and $P_{tx}+P_{ty}$ transform as the $A_1^+$
(scalar) component of the PV and have non-vanishing expectation values
at finite temperature. On the other hand, $P_{tx}-P_{ty}$ transforms
as the $B_1^+$ \cite{glue}. Due to the reasons given earlier, this last
expectation value must vanish, and we show later that it does. However,
the correlation function need not, and, indeed, does not.
Non-trivial screening
has been observed through gauge-invariant gluonic correlation functions
\cite{glue} in all the other quantum number channels (labelled by the
irreps of $D_4^h$) as well. 

Screening masses obtained from correlation functions built out of
staggered Fermion field operators have also been extensively studied in the
past \cite{screen,mtc,myscr}. The correlators which have been measured
in the past are the $A_1^+$ from the scalar (S) and pseudo-scalar (PS)
channels, and $A_1^+$ combinations of the vector (V) and pseudo-vector
(PV) channels. The two $A_1^+$ correlators descending from S and PS
see a lower screening mass ($\mu$) than those descending from the V and
PV. The latter are consistent with the expectation from free Fermion
field theory---
\beq
   \mu a=2\sinh^{-1}\left(\sqrt{(ma)^2
               +\sin^2\left(\frac\pi{N_t}\right)}\right),
\label{fft}\eeq
where $a$ is the lattice spacing, $m$ the quark mass, and $N_t$ is the
number of lattice sites in the Euclidean time direction ($T=1/N_ta$).
Even some other measurements, such as those of ``wavefunctions'', 
which seemed to indicate a more complicated picture \cite{milc}, 
can be understood in terms of weakly interacting quarks \cite{suny}. 
Here we re-examine the  screening masses with the complete decomposition of 
operators into irreps of the finite temperature invariance group. In
particular, we report in this letter the results of the first measurement of 
the $B_1^+$ correlation function constructed from local Fermion bilinears
(see Table III of \cite{fermit})
and compare our results with those obtained with gluonic operators. 

We have simulated QCD with four light degenerate flavors of quarks at
temperatures above the phase transition temperature, $T_c$, on lattices
of size $4\times10^2\times16$, using the Hybrid Monte Carlo (HMC)
algorithm \cite{hmcalg}. The longest direction, $N_z$, was chosen to
be four times the Euclidean time direction, $N_t$, so that correlations
could be followed to a distance of $2/T$. One simulation was performed at
$T=3T_c/2$ with the coupling $\beta=5.1$ and the quark mass $m=0.015/a$
where $a$ is the lattice spacing. The second simulation was made at
$T=2T_c$ with $\beta=5.15$ and $m=0.01/a$. With our choice of $N_t =
4$, $a = 1/4T$. The temperature identifications are made using previous
measurements of the critical coupling on lattices with larger values of
$N_t$ \cite{crit}. Companion runs were made in quenched QCD on lattices
of the same size at couplings corresponding to $3T_c/2$ and $2T_c$ for
the quenched theory using a Cabbibo-Marinari pseudo-heat-bath algorithm
\cite{pheat}.

We thermalised the HMC simulation at $3T_c/2$ with two different runs--- one
starting from an ordered gauge configuration, and the other from a pure
gauge configuration thermalised at $2T_c$. Agreement in measurements of
all thermodynamic quantities was used to decide on thermalisation. The
plaquette average turned out to be the most stringent test, since it is
the least noisy. At $2T_c$ thermalisation was tested by checking that
a run starting from an ordered gauge configuration gave the same
thermodynamics as one starting from a thermal $3T_c/2$ configuration.

Once thermalisation was achieved, two runs were made at each
temperature--- one with a trajectory length of one molecular dynamics
(MD) time unit, and another with a trajectory length half as long. At
$3T_c/2$ statistics were collected from 875 such configurations
generated using the long trajectory, and 285 with the short trajectory.
Previous studies have shown that the physics is much simpler at $2T_c$. 
We did smaller runs at this temperature--- collecting statistics of 445
configurations with the short trajectory length and 100 with the long
trajectories.

The question of autocorrelations is important whenever statistical
inferences are to be made. It was found that autocorrelations of local
operators, such as plaquettes, were the same with the two different
trajectory lengths mentioned above. However, with any simulation algorithm 
that undergoes critical slowing down, short distance operators are
decorrelated faster than those which are dominated by large distance
scales. Thus, the effective number of measurements of short distance
operators is not the same as that for extended operators. This is most
problematic for correlation function measurements, where the correlator
at different distances may have entirely different autocorrelations.
These are usually difficult to measure directly and a different
approach to the problem seems necessary.

If the errors in Fermion bilinear correlation functions, $\Delta C(z)$,
are evaluated with the assumption that there are no autocorrelations,
then they depend systematically on the separation $z$. We found that
for sufficiently large statistics, $\Delta C(z)$ falls exponentially 
with $z$ and its logarithmic slope is almost independent of statistics.
Hence it is possible to quote a single number as a figure of merit for
decorrelation---
\beq
   D=\frac{\Delta C(N_z/2)}{\Delta C(0)}.
\label{merit}\eeq
$D$ is usually larger than unity, and depends on the specifics of the
simulation algorithm and its tuning. Since smaller values of $D$ are
preferable, tuning the algorithm should be done to minimise this.

The values of $D$ obtained depended on the channel being studied: the
largest values of $D$ were found in the $A_1^+$ irreps coming from the
V or the PV, and the smallest in the $B_1^+$ irreps.  In each channel,
we found only a weak dependence of $D$ on the number of measurements---
indicating that it is a direct measure of the efficiency of the algorithm.

We found that the long trajectories (1 MD time unit) give about half
the value of $D$ as obtained with the shorter trajectory. Since it
takes twice as long to run the longer trajectory, the computational
effort,
 $ E = D\times(CPU {\rm\ time})$, 
involved in getting equal statistical errors is the same for these
two trajectory lengths. However, the analysis of correlated errors in
screening mass measurements is simplified with the longer trajectories.
In test runs with trajectories 2.5 MD time units long, we found no
further decrease in $D$, and hence an increase in $E$. Such a dependence
of $E$ on trajectory length is characteristic of the
HMC simulation algorithm \cite{hmc}. In the dynamical QCD simulations,
we found $D$ to lie in the range 250 to 1000.  The quenched simulations
were significantly easier to decorrelate--- the value of $D$ was lower
by a factor of roughly 40 compared to the full theory simulations. This
information on autocorrelations has been incorporated in all our
statistical analyses.

In Table \ref{tb.masses}, we report our measurements of screening
masses for the known $A_1^+$ correlators. Our results are in very good
agreement with previous measurements in the quenched theory at $N_t=4$
\cite{myscr}. There is also a remarkable agreement between the $A_1^+$
screening masses obtained from the quenched and the dynamical Fermions
simulations at both temperatures. The four $A_1^+$ irreps coming from
the V and PV channels gave degenerate screening masses which agree
extremely well with the free field theory estimate in eq.\ (\ref{fft}).
The $A_1^+$ correlators in the S and PS channels gave smaller screening
masses, which increase marginally with $T$.

Hadron mass measurements in 4-flavor QCD at quark mass, $ma=0.01$ and
$\beta=5.15$, corresponding to our runs at $2T_c$, have been performed
before \cite{edwin}. A comparison with these $T=0$ measurements, listed
in the last column of Table \ref{tb.masses}, shows that our finite
temperature screening masses are completely different---
\beq
   \frac{\mu(T)}{m(T=0)}\approx
     \cases{3&\qquad($A_1^+$ from PS),\cr
            2&\qquad($A_1^+$ from V).}
\eeq
In contrast, earlier measurements for $N_t=8$ lattices yielded
$\mu/m\approx 1$ for the $A_1^+$ screening mass in the vector channel
V  \cite{mtc}. This made it difficult to argue for a thermal effect,
although $\mu$ agreed with eq.\ (\ref{fft}) even in that case. The
present measurement resolves this problem.

\begin{figure}[htbp]\begin{center}
   \leavevmode
   \psfig{figure=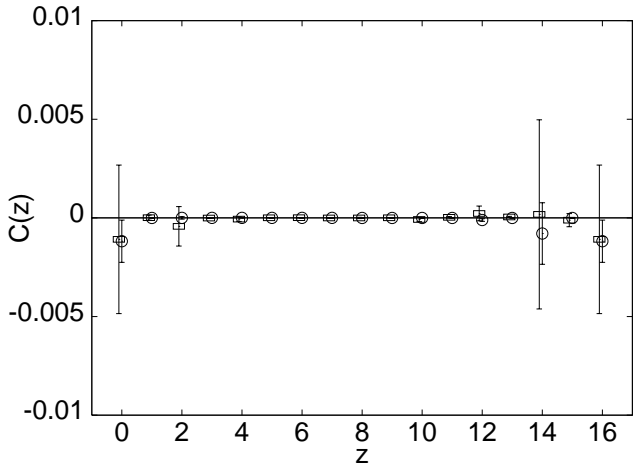,height=6cm}
   \end{center}
   \caption{The $B_1^+$ correlation function obtained from the V channel
       operators. Boxes denote data for dynamical QCD simulations for
$3T_c/2$ and circles for $2T_c$.}
\label{fg.b1p}\end{figure}

Our main new result is the first measurement of the correlation function
in the $B_1^+$ channels. Correlators in this irrep obtained with PV and
V are identical up to a sign, configuration by configuration. Hence
we restrict our attention only to the $B_1^+$ coming from the V. We
found that the $B_1^+$ correlation functions vanish to the best of the
measurement precision (see Figure \ref{fg.b1p}). The correct statistical
procedure is to quote the $\chi^2$ value for the fit of the data by a
correlation function which is identically zero. We found
\beq
   \chi^2/{\rm DOF} = \cases{7/15 & ($\frac{3}{2}T_c$, dynamical),\cr
                             11/15 & ($2T_c$, dynamical).\cr}
\label{chisqd}\eeq
The quenched runs gave very similar results---
\beq
   \chi^2/{\rm DOF} = \cases{7/15 & ($\frac{3}{2}T_c$, quenched),\cr
                             5/15 & ($2T_c$, quenched).\cr}
\label{chisqq}\eeq
The numbers in eq.\ (\ref{chisqd}) have been obtained with the statistics
collected in the longer of the two runs made at each temperature. The
HMC simulations with smaller statistics also gave very similar results
at both these temperatures.

One possible explanation for these remarkable results is 
an exact symmetry between the $x$ and
$y$ directions, configuration by configuration. If this were so, then
non-scalar operators would vanish, not only on the average, but
identically.  As a result, the correlation function in all but
the scalar channel would also vanish.

We test for this symmetry by correlation functions of the $B_1^+$ plaquette
operator, discussed earlier. Its average must vanish, and does---
\beq
\langle  P_{tx}-P_{ty} \rangle = \cases{
(-0.6 \pm 1.8) \times 10^{-4} & ($\frac{3}{2}T_c$, quenched),\cr
(5.5 \pm 3.8) \times 10^{-5} & ($2T_c$, quenched).\cr}
\label{plaq}\eeq
However, the corresponding correlation function does not vanish.  The
same configurations used in the Fermionic correlator measurements lead to
gluonic $B_1^+$ correlations which are clearly non-zero. 

\begin{figure}[htbp]\begin{center}
   \leavevmode
   \psfig{figure=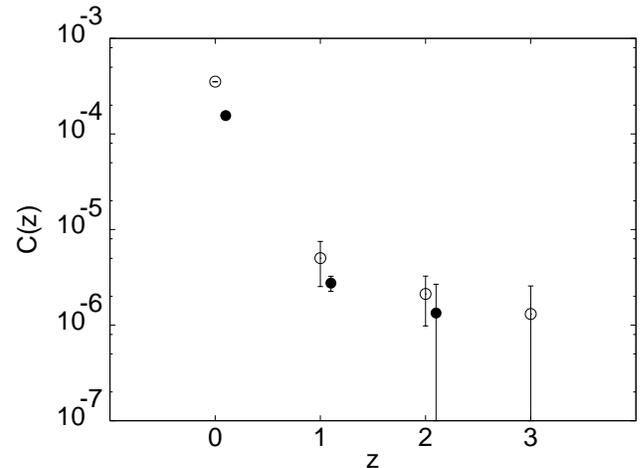,height=6cm}
   \end{center}
   \caption{The $B_1^+$ correlation function obtained from the 
       plaquette operators. Filled circles denote data for $3T_c/2$ and open 
       for 2$T_c$. }
\label{fg.b1g}\end{figure}

This correlation function is exhibited in Figure \ref{fg.b1g}
for the quenched QCD simulations. Since the corresponding screening
mass is large \cite{glue}, the correlation function is somewhat noisy
at large distances but it is clearly very different from zero at small
distances. The test of a vanishing value of this correlation function
gives $\chi^2/{\rm DOF}\approx800$.  Thus the gauge configurations
underlying our measurements have a strong configuration-to-configuration
asymmetry between the $x$ and $y$ directions, allowing the gluonic
$B_1^+$ correlators to survive.

Another possibility is that we have accidentally chosen an operator 
which has small overlap with
the lowest $B_1^+$ eigenvector of the transfer matrix.  One way to test
this is to work with other operators. It is often the case that the overlap
on the lowest eigenvector improves by delocalising the operator in some
way. We constructed the ``meson'' operators using quark propagators
computed with fuzzed links. This fails to improve the $B_1^+$
correlation function. We conclude that there must be a dynamical reason
for the vanishing of the $B_1^+$ correlator in the high temperature
phase of QCD, when measured using Fermion bilinears.

The numerical agreement of the screening masses extracted from the
exponential fall-off of the $A_1^+$-correlators with free field values
of eq.\ (\ref{fft}) has been used earlier to argue that one sees weakly
interacting quarks in the high temperature phase of QCD.  In the free-field
theory limit, the Fermionic $B_1^+$ correlators studied here would
vanish.  Consequently, our observations can be seen as additional evidence
for the weakly interacting picture. Indeed, comparing the 
$\chi^2$ values, we find that the $B_1^+$ effective 
coupling in the quark sector is approximately 40 times smaller than that in the 
gluon sector, assuming the overlaps to be similar.
Since the vanishing or existence of
a correlation function is easy to observe, we believe that the $B_1^+$
correlator is a qualitatively better indicator of the non-interacting
nature of the quarks in the quark-gluon plasma.

The drawback of this picture of non-interacting Fermions is well-known---
the $A_1^+$ irreps coming from the S and PS channels are not degenerate
with those coming from the V and PV channels. A plausible argument to
understand this phenomenon is to note that the $A_1^+$ correlator coming
from the S channel mixes with the glue sector of the theory. As a result,
the screening mass in this channel will be contaminated by those in
the gluonic $A_1^+$ sector, of which the lowest is the Debye screening
mass, $m_D$. In quenched simulations at $3T_c/2$, $m_D/T = 2.8 \pm 0.2 $
\cite{glue}. Since the screening mass for the S channel lies in between
$m_D$ and the V channel mass, as seen in Table \ref{tb.masses}, it is
consistent with such a conjecture.

We employed staggered Fermions for this investigation. It would be
interesting to confirm these results for the Wilson Fermions as well.
The symmetries of the lattice are, of course, independent of the type
of Fermions employed and the group theoretic arguments apply with small
modifications. In particular, the break-up of the zero temperature
spectrum under the $D_4^h$ group proceeds without change, although the
actual operators realising the irreps do change.

To summarize, we have used a new Fermionic correlator to demonstrate
that the quarks in the quark-gluon plasma are weakly interacting already
at temperatures as low as $3T_c/2$. The particular correlator we used
is a much better probe of Fermion interaction strength than those used
earlier.


\begin{table}[htbp]\begin{center}
\begin{tabular}{cccc}  \hline
Channel & $3T_c/2$ & $2T_c$ & $T=0$ \\
\hline
S $A_1^+$  & $0.91(3)$ & $1.07(4)$ & $0.60(5)$\cr
PS $A_1^+$ & $0.91(3)$ & $1.07(4)$ & $0.303(2)$\cr
V $A_1^+$  & $1.39(7)$ & $1.35(6)$ & $0.71(6)$\cr
PV $A_1^+$ & $1.39(7)$ & $1.35(6)$ & $1.4(1)$\cr
\hline
\end{tabular}\end{center}
\caption[dummy]{Screening masses in units of the lattice cutoff, $1/a
    = 4T$. The $T=0$ masses quoted here were measured \cite{edwin}
    with the same lattice spacing and quark mass as the runs at
    $2T_c$.}

\label{tb.masses}\end{table}

\end{multicols}

\end{document}